\documentclass[letterpaper]{article}
\pdfoutput=1
\usepackage{jheppub}
\usepackage{amsmath}
\usepackage{graphicx}
\usepackage[config=altsf]{subfig}
\usepackage{feynmf}
\usepackage{dsfont}
\usepackage{array}
\usepackage{slashed}
\usepackage{afterpage}
\usepackage{appendix}
\usepackage{multirow}

\newcommand{\ie}{{\it i.e.}}

\usepackage[usenames,dvipsnames]{xcolor}

\usepackage{dcolumn}
\usepackage{slashed}

\newcommand{\be}{\begin{equation}}
\newcommand{\ee}{\end{equation}}
\newcommand{\bea}{\begin{eqnarray}}
\newcommand{\eea}{\end{eqnarray}}
\newcommand{\lsim}{\!\mathrel{\hbox{\rlap{\lower.55ex \hbox{$\sim$}} \kern-.34em \raise.4ex \hbox{$<$}}}}
\newcommand{\gsim}{\!\mathrel{\hbox{\rlap{\lower.55ex \hbox{$\sim$}} \kern-.34em \raise.4ex \hbox{$>$}}}}

\def\eV{\text{eV}}

\title{Cosmology in Mirror Twin Higgs and Neutrino Masses}
\author[a]{Zackaria Chacko,}
\author[b]{Nathaniel Craig,}
\author[c]{Patrick J. Fox}
\author[c]{and Roni Harnik}
\affiliation[a]{Maryland Center for Fundamental Physics,
Department of Physics, \\
University of Maryland, College Park, MD 20742, USA}
\affiliation[b]{Department of Physics, University of California, Santa Barbara, 
CA 93106, USA}
\affiliation[c]{Fermilab, P.O. Box 500, Batavia, IL 60510, USA}

\emailAdd{zchacko@physics.umd.edu}
\emailAdd{ncraig@physics.ucsb.edu}
\emailAdd{pjfox@fnal.gov}
\emailAdd{roni@fnal.gov}

\abstract{
We explore a simple solution to the cosmological challenges of the 
original Mirror Twin Higgs (MTH) model that leads to interesting 
implications for experiment. We consider theories in which both the 
standard model and mirror neutrinos acquire masses through the familiar 
seesaw mechanism, but with a low right-handed neutrino mass scale of 
order a few GeV. In these $\nu$MTH models, the right-handed neutrinos 
leave the thermal bath while still relativistic. As the universe 
expands, these particles eventually become nonrelativistic, and come to 
dominate the energy density of the universe before decaying. Decays to 
standard model states are preferred, with the result that the visible 
sector is left at a higher temperature than the twin sector. 
Consequently the contribution of the twin sector to the radiation 
density in the early universe is suppressed, allowing the current bounds 
on this scenario to be satisfied. However, the energy density in twin 
radiation remains large enough to be discovered in future cosmic microwave 
background experiments. In addition, the twin neutrinos are 
significantly heavier than their standard model counterparts, resulting 
in a sizable contribution to the overall mass density in neutrinos that 
can be detected in upcoming experiments designed to probe the large 
scale structure of the universe.
}

\preprint{FERMILAB-PUB-16-555-T\\
November 23, 2016
}

\begin{document}


\maketitle

\section{Introduction}

Models that address the hierarchy problem using a symmetry to protect 
the Higgs mass predict the existence of top partners near the 
electroweak scale. In most familiar realizations of such a symmetry, 
such as supersymmetry~\cite{Fayet:1977yc, Dimopoulos:1981zb} or Little Higgs models~\cite{ArkaniHamed:2001nc}, the top 
partners are colored, and would therefore be produced at the Large 
Hadron Collider (LHC) with high rates. The absence of evidence for such 
particles has led to increasing interest in models based on the 
framework of Neutral 
Naturalness~\cite{Chacko:2005pe,Barbieri:2005ri,Chacko:2005vw,Burdman:2006tz,Cai:2008au,Poland:2008ev,Batell:2015aha}. 
In this class of theories the top partners are not colored, thereby 
providing a natural explanation for their elusiveness.

The earliest and perhaps most elegant realization of Neutral Naturalness 
is the Mirror Twin Higgs (MTH) model~\cite{Chacko:2005pe}, in which a 
mirror (``twin") copy of the standard model (SM) is introduced. A 
discrete $\mathbb{Z}_2$ symmetry relates the particle content and interactions of 
the SM and twin sectors. Although this $\mathbb{Z}_2$ symmetry is not exact, it 
is only broken softly. This allows the Twin Higgs vev $f$ to be a factor 
of a few larger than the electroweak vev $v$. In this theory the top 
partners, and, for that matter, all BSM particles with masses below a 
TeV, are SM singlets. The cancellation of divergences in the Higgs 
potential is realized through a ``Higgs Portal" coupling of the SM Higgs 
to the Twin Higgs. The LHC signals of the mirror symmetric twin 
framework include modified Higgs couplings as well as an invisible 
branching fraction~\cite{Burdman:2014zta}.

Since the original proposal, the Twin Higgs framework has been further 
developed. Several possible UV completions of the MTH model have been 
proposed - 
supersymmetric~\cite{Falkowski:2006qq,Chang:2006ra,Craig:2013fga}, 
holographic~\cite{Geller:2014kta}, 
composite~\cite{Barbieri:2015lqa,Low:2015nqa}, and extra 
dimensional~\cite{Craig:2014aea,Craig:2014roa}. Phenomenological and 
model-building aspects of the framework and its UV completions have also 
been explored. Flavor constraints on the Holograpic and composite Twin 
Higgs have recently been presented finding reduced tension as compared 
to regular composite Higgs models~\cite{Csaki:2015gfd}.  Various options 
for the breaking of $\mathbb{Z}_2$ and the way they affect the tuning of 
the model have also been investigated~\cite{Beauchesne:2015lva, 
Harnik:2016koz}.

The cosmology of the MTH model is somewhat problematic. The Higgs portal 
interaction maintains thermal equilibrium between the SM and its twin 
copy down to temperatures of order a few GeV~\cite{Barbieri:2005ri}. 
Below this temperature the twin sector continues to have a sizable 
contribution to the total energy density in radiation, from mirror 
photons and neutrinos at late times. This brings the theory into 
conflict with the tight constraints on dark radiation from Big Bang 
Nucleosynthesis (BBN), and from the Cosmic Microwave Background (CMB). 
This limit on dark radiation is often quoted as a limit on the effective 
additional number of neutrinos $\Delta N_\mathrm{eff}$. A small hard 
breaking of the $\mathbb{Z}_2$ symmetry in the Yukawa sector, as 
discussed in~\cite{Barbieri:2016zxn}, offers a minimal approach to 
address the cosmological problems of the MTH, although here we shall 
follow a different route. Once the bounds on $\Delta N_\mathrm{eff}$ are 
satisfied, cosmological puzzles such as the origin of dark 
matter~\cite{Farina:2015uea} and the generation of the baryon 
asymmetry~\cite{Farina:2016ndq} can be addressed. We note that 
conventional mirror models~\cite{Foot:1991py, Foot:1991bp, 
Berezhiani:1995am} are able to avoid the bounds on $\Delta 
N_\mathrm{eff}$ by simply eliminating the Higgs portal coupling between 
the two sectors. However, in Twin Higgs constructions this coupling 
plays a critical role in the cancellation of quadratic divergences, and 
so this is not a viable option.

Recently, alternative realizations of the Twin Higgs have been proposed 
in which the twin particle content is smaller than in the MTH, 
consisting only of those states needed to address the naturalness 
problem. This Fraternal Twin Higgs (FTH) framework~\cite{Craig:2015pha} 
includes only the third generation of fermions, the electroweak gauge 
bosons and the twin gluon. As a consequence of the reduced particle 
content, this class of theories is free of the cosmological challenges 
of the MTH model. FTH models also predict new and interesting LHC 
signals since the lightest twin particles, the glueballs, will be 
produced in Higgs decays and will naturally decay displaced from the 
interaction point. This opens a new opportunity for LHC to probe these 
scenarios~\cite{Curtin:2015fna, Csaki:2015fba, Chou:2016lxi}. Dark 
matter can be naturally accommodated within this 
framework~\cite{Craig:2015xla,Garcia:2015loa,Garcia:2015toa,Freytsis:2016dgf}. 
It can also be used to explain certain anomalies in large and small 
scale structure~\cite{Prilepina:2016rlq}. Within this construction the 
twin sector can be vector-like~\cite{Craig:2016kue}, removing the need 
for the third generation twin leptons, which would otherwise be needed 
for anomaly cancelation.

In this paper we show that a very simple extension of the original MTH, 
without any additional breaking of the discrete $Z_2$ twin symmetry, can 
evade these cosmological difficulties. In this framework a new particle 
species $N$ decouples from the thermal bath while still relativistic, 
and comes to dominate the energy density of the universe at late times. 
The decay of the $N$ is preferentially to SM particles, thus heating our 
sector and effectively diluting the energy density of the twin sector.  
Consequently the contribution of the mirror sector to the radiation 
density in the early universe is suppressed, allowing the current 
cosmological bounds on dark radiation to be satisfied. However, the 
contribution of the twin sector to $\Delta N_\mathrm{eff}$ is in general 
large enough to be observed in future CMB experiments such as 
SPT-3G~\cite{Benson:2014qhw} and ACT~\cite{Naess:2014wtr}.

This mechanism arises naturally in the $\nu$MTH, a minimal extension of 
the MTH that incorporates neutrino masses. In the $\nu$MTH, both the SM 
and mirror neutrinos acquire masses through the Type-I seesaw mechanism, 
but with a low right-handed neutrino mass scale of order a few GeV. In 
this scenario, it is the right-handed neutrinos that play the role of 
the $N$. They decouple from the SM bath while still relativistic. As the 
universe expands they redshift, become nonrelativistic, and eventually 
come to dominate the energy density of the universe before decaying. 
Even if the neutrino sector fully respects the $\mathbb{Z}_2$ symmetry, 
decays to SM states are preferred because of the hierarchy of 
electroweak vevs $f > v$, with the result that the visible sector is 
left at a higher temperature than the mirror sector\footnote{Another 
framework for addressing the hierarchy problem in which a new particle 
decays preferentially into the sector with the lightest electroweak vev 
was presented in~\cite{Arkani-Hamed:2016rle}, though both the number of 
sectors and assumptions about decoupling are different.}.  In this model 
the mirror neutrinos are significantly heavier than their SM 
counterparts, resulting in a sizable contribution to the overall 
cosmological mass density in neutrinos that can be detected by future 
probes of large scale structure such as DES~\cite{Lahav:2009zr}, 
LSST~\cite{Abell:2009aa} and DESI~\cite{Levi:2013gra}. This framework 
for neutrino masses therefore offers a natural resolution to the 
cosmological problems of the original proposal, while leading to 
interesting predictions for upcoming experiments.

The outline of this paper is as follows. In the next section, we discuss 
in greater detail the cosmological problems of the original MTH model. 
In Section \ref{sec:cosmology} we explore the range of parameter space 
in which a long-lived massive particle that decays preferentially into 
visible sector states can give rise to a sufficiently small $\Delta 
N_\mathrm{eff}$. In Sections \ref{sec:neutrinomodel} and 
\ref{sec:threefam} we introduce the $\nu$MTH model, in which neutrino 
masses are incorporated into the Twin Higgs framework via a Type-1 
seesaw, and show that there is a range of parameter space in which the 
late decays of right-handed neutrinos can solve the cosmological 
problems of the original MTH model.  We conclude in Section 
\ref{sec:conclusions}.

\section{Cosmology in the Mirror Twin Higgs}
\label{sec:problem}

As mentioned above, the original MTH model predicts an abundance of dark 
radiation in the early Universe, in conflict with observation. In this 
section we review the problem and assess its severity. Following the 
established convention, we use the label $A$ to denote visible sector 
states, and the label $B$ to denote twin sector states.

\begin{figure}[t] 
   \centering
   \includegraphics[width=2in]{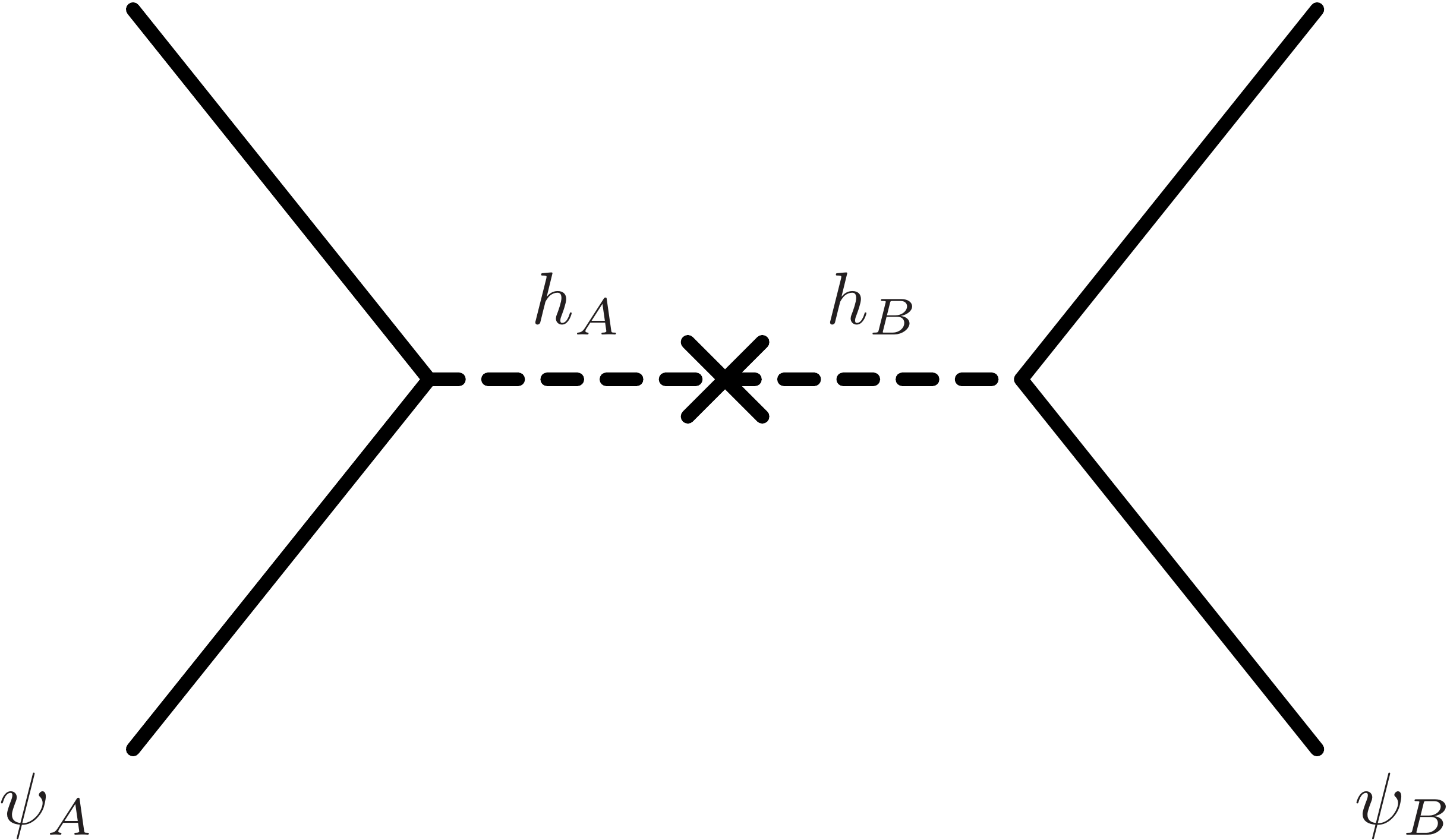} 
   \caption{Higgs portal interaction that keeps the $A$ and $B$ sectors in equilibrium. }
   \label{fig:ABequilibrium}
\end{figure}

In the original MTH model, the $\mathbb{Z}_2$ symmetry is explicitly broken, but 
only softly. As a result of this soft breaking, the vev of the SM Higgs 
$\langle H_A\rangle =v=246$~GeV is smaller than the vev of the twin 
Higgs $\langle H_B \rangle = f$ by a factor of a few. In the MTH 
framework, the cancellation of quadratic divergences arises from a Higgs 
portal interaction between the SM Higgs doublet and its twin partner. As 
a consequence of this interaction, after electroweak symmetry breaking 
the SM Higgs $h_A$ and its twin partner $h_B$ mix, so that the lightest 
Higgs state $h$ is a linear combination of these two states,
 \begin{equation}
h\sim \cos\left(\frac{v}{f}\right) h_A +  \sin\left(\frac{v}{f}\right)h_B.
\end{equation}
 The state $h$ is identified with the Higgs boson that has been observed 
with mass 125 GeV. Each of the Higgs bosons $h_A$ and $h_B$ only has 
Yukawa couplings to the fermions in its own copy of the standard model. 
Consequently the mass eigenstate $h$ will couple to both sets of 
fermions, but with an interaction strength suppressed by the mixing. As 
a result, in the early universe the 125~GeV Higgs mediates the 
scattering of $A$ and $B$ femions off one another, see 
Fig~\ref{fig:ABequilibrium}. This leads to an interaction rate between 
the two sectors of order
 \begin{equation}
\langle \sigma v \rangle \simeq \left(y_A^i y_B^j\right)^2 \frac{v^2}{f^2} \frac{T^2}{m_h^4}.
 \end{equation}
 Here $T$ denotes the temperature of the bath, and $y_A^i$ and $y_B^j$ 
represent the Yukawa couplings of the heaviest fermion that is in 
equilibrium in the corresponding sector at that temperature.  
Equilibrium between the $A$ and $B$ sectors is maintained down to the 
temperature at which the scattering rate is comparable to the Hubble 
expansion rate, $n\langle \sigma v \rangle \sim H$. Applying this 
formula, we find that the Higgs portal interaction keeps the $A$ and $B$ 
sectors in equilibrium down to a temperature $T_D$ of order 3 GeV. Below 
this temperature the two sectors decouple. It should be noted that this 
Higgs portal interaction is an integral part of the twin mechanism which 
addresses the hierarchy problem in this framework. For this reason it is 
not possible to reduce the ratio $v/f$ significantly without introducing 
an unacceptable amount of tuning into the theory.

Let us now estimate the energy density in mirror radiation. It will be 
particularly convenient to focus on the ratio of energy densities of the 
two sectors, $\rho_B/\rho_A$. At the decoupling temperature $T_D$ the 
temperatures of the two sectors are equal, and therefore the ratio of 
energy densities of the two sectors is simply the ratio of the effective 
number of degrees of freedom,
 \begin{equation}
T=T_D \sim\mbox{GeV :} \qquad \left.\frac{\rho_B}{\rho_A}\right|_{T_D} 
= \left.\frac{g_{*B}}{g_{*A}}\right|_{T_D} 
\label{dofTD}
 \end{equation} 
 The effective number of degrees of freedom in each sector is defined 
in the usual way,
 \begin{equation}
g_*=\sum \left( N_\mathrm{bosons} + \frac{7}{8} N_\mathrm{fermions} \right) \; ,
 \end{equation}
 where the sum is over the degrees of freedom in each sector which are 
in equilibrium at a particular temperature. For concreteness, we choose 
to evaluate the $g_*$ at 3 GeV. In the SM sector at that temperature 
we include all fermions with masses between those of the electron and 
the tau, as well as three generations of neutrinos, the gluons and the 
photon. In the twin sectors we include the same matter content, but 
without the mirror charm and tau. This gives us
 \begin{equation}
\left. g_{*A}\right|_{T=\mathrm{3\, GeV}} = \frac{303}{4} \qquad \mbox{and} \qquad
\left. g_{*B}\right|_{T=\mathrm{3\, GeV}} = \frac{247}{4}\,.
 \end{equation}
 Applying this to Eq.~(\ref{dofTD}), we find that this translates to an 
energy density ratio of about 0.8 when the two sectors decouple.

The bounds on energy density in hidden radiation come from Big Bang 
Nucleosynthesis (BBN), and from the Cosmic Microwave Background (CMB), 
so $\rho_B$ needs to be evaluated at these later times. To calculate the 
ratio of energy densities it is therefore necessary to account for the 
degrees of freedom that have left the bath, become nonrelativistic, and 
annihilated in both sectors. As species leave the thermal bath the 
comoving entropy is conserved. As a result, when a sector transitions 
from an initial effective number of degrees of freedom $g_*^{(i)}$ to a 
lower number $g_*^{(f)}$, the comoving energy density increases by a 
factor of $(g_*^{(i)}/g_*^{(f)})^{1/3}$. By the time of BBN (or CMB) all 
species decouplings have already occurred, and so the effective $g_*$ is 
identical in both the $A$ and $B$ sectors, $g_{*A}^\mathrm{BBN}= 
g_{*B}^\mathrm{BBN}$\footnote{One also needs to consider that 
electron-positron decoupling happens after neutrino decoupling making 
the effective increase in energy density differ from the naive formula 
above. However, neutrino decoupling precedes electron annihilation in 
both sectors and the correction to energy densities in both sectors is 
identical.}. As a result $g_*^{(f)}$ is the same in both sectors. Then 
the ratio of energy densities at late times, including the corrections 
from species leaving the baths, is given by
 \begin{equation}
 \left.\frac{\rho_B}{\rho_A}\right|_\mathrm{BBN} \simeq
 \left(\left.\frac{g_{*B}}{g_{*A}}\right|_{T_D}\right)^{1/3} 
\left.\frac{\rho_B}{\rho_A}\right|_{T_D} =  
 \left(\left.\frac{g_{*B}}{g_{*A}}\right|_{T_D}\right)^{4/3}
\approx 0.75
\end{equation}
 
 We are now in a position to determine the corrections to $\Delta 
N_\mathrm{eff}$ from the twin sector. At late times, after the neutrinos 
have decoupled and the positrons have left the bath, neutrinos make up 
roughly 0.4 of the energy density in SM radiation. It is then 
straightforward to translate a limit on $\Delta N_\mathrm{eff}$ into a 
limit on the ratio of energy densities in the $A$ and $B$ sectors,
 \begin{equation}
\Delta N_\mathrm{eff} = 
3 \left.  \frac{\rho_B}{\rho_{\nu}} \right|_{\mathrm{BBN}}
\approx 7.4 \left.\frac{\rho_B}{\rho_{A}}\right|_{\mathrm{BBN}} \approx 5.6 \; .
\label{eq:neffatMeV}
\end{equation} 
 Here $\rho_\nu$ refers to the energy density in all SM neutrinos. 
Current bounds on $\Delta N_\mathrm{eff}$ from BBN are of order 0.5-1, 
depending on the input~(see for example~\cite{Cyburt:2004yc}). The 
limits on $\Delta N_\mathrm{eff}$ from the CMB are more stringent, and 
require $\Delta N_\mathrm{eff} \lesssim 0.6$ at $2 
\sigma$~\cite{Ade:2015xua}. This bound may be somewhat relaxed if the 
dark radiation scatters with a short mean free path, as opposed to free 
streaming like neutrinos~\cite{Baumann:2015rya}. It should also be noted 
that the recently observed tension between CMB observations and the 
measurement of the local Hubble expansion can be interpreted as a hint 
of a $\Delta N_\mathrm{eff}\sim O(1)$~\cite{Riess:2016jrr}. However, even after 
taking these factors into account, it is clear that an energy density 
ratio $\rho_A/\rho_B$ as large as predicted by the original MTH 
model (\ref{eq:neffatMeV}), is ruled out both by Big Bang Nucleosynthesis (BBN), and the 
Cosmic Microwave Background (CMB).

This tension with cosmology can be addressed in several ways. For 
example, if the two sectors were to decouple at a time when there are 
significantly more degrees of freedom in the visible sector than in the 
twin sector, the ratio of energy densities would decrease, leading to a 
smaller $\Delta N_\mathrm{eff}$. This was considered in \cite{Farina:2015uea}, which 
explored a scenario in which decoupling occurred after the QCD phase 
transition in the twin sector, but before that in the visible sector. 
However, the effective number of degrees of freedom in the two sectors 
does not differ sufficiently, even during this short epoch, to fully solve 
the problem. The model in \cite{Farina:2015uea} can therefore accommodate a photon in 
the twin sector, but not the mirror neutrinos.

Another possibility is to simply remove all of the ``unnecessary'' light 
degrees of freedom from the twin sector. In the FTH 
model~\cite{Craig:2015pha} the twin sector is taken to contain only the 
third generation of fermions, as well as the twin EW and QCD gauge 
bosons. One can further remove light degrees of freedom by assuming that 
the twin sector is vector-like, as in~\cite{Craig:2016kue}. This removes 
the need for the twin tau neutrino.

\section{A Viable Cosmology: Matter Domination and Preferential Decays}
\label{sec:cosmology}

In this paper we will focus on the MTH framework, in which the full 
matter content of the SM is replicated in the twin SM, with identical 
Yukawa couplings. The twin sector then contains three light neutrinos 
and a massless photon. Although these light twin states will be 
thermalized in the early universe, we now show that by minimally 
extending the original MTH model, their contribution to the energy 
density at late times can be suppressed.

In the early universe the SM and twin SM are kept in thermal equilibrium 
through interactions mediated by the Higgs portal. When the two sectors 
decouple, which happens at a temperature $T_D$ of order a few GeV, the 
SM and its twin are at the same temperature and contain roughly the same 
number of degrees of freedom. To realize our scenario we introduce into 
the theory one or more new particles $N$ that lie outside the SM, and 
have masses $M_N$ above a GeV. These new particles are assumed to have 
very small couplings to the SM, and therefore decouple from the SM bath 
while still relativistic. They then survive for a time as thermal 
relics, become nonrelativistic, and eventually come to dominate the 
energy density of the universe before decaying. If these decays are 
preferentially to SM states rather than to twin states, and furthermore 
occur after the the two sectors have decoupled, the SM will be left at a 
higher temperature than its twin counterpart. Consequently the energy 
density of the SM sector will be larger than that of the twin sector, 
allowing the bounds on $\Delta N_{\rm eff}$ to be satisfied.

 The name $N$ is chosen in anticipation of the identification on these 
new particles as right-handed neutrinos which, as we shall see, can 
easily satisfy these requirements. We note, however, that the $N$ 
could be be identified with any particles in the theory that are 
sufficiently heavy and sufficiently long lived. To emphasize this, we 
will remain agnostic about the identity of the $N$ in this section. We shall 
simply parametrize the framework in terms of their mass $m_N$, their width 
$\Gamma_N$, the effective number of degrees of freedom in the $N$ sector 
$g_{*N}$ and the fraction of decays into hidden sector states $\epsilon$,
 \begin{equation}
\epsilon \equiv \frac{\Gamma_{N\to B}}{\Gamma_N} \ll 1 .
 \end{equation}
 These parameters are sufficient to compute the energy density in twin 
radiation, which can then be translated into~$N_\mathrm{eff}$. 

\begin{figure}[t] 
   \centering
   \includegraphics[width=0.6\columnwidth]{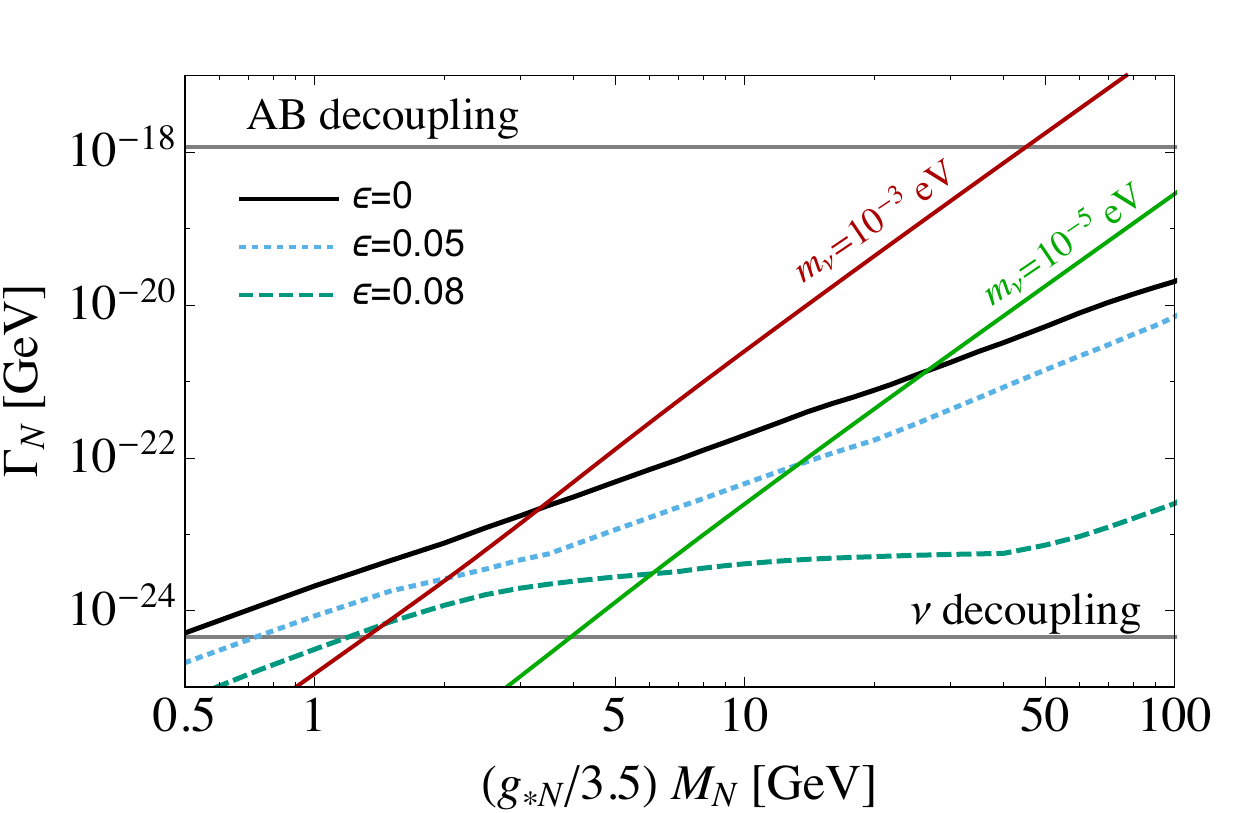} 
  \caption{The constraints on $m_N-\Gamma_N$ parameter space. The reheat temperature after $N$ decay must lie above the neutrino decoupling temperature, taken to be 1 MeV, and below the SM-Twin decoupling temperature, taken to be 3 GeV.  We presents curves of $\Delta N_\mathrm{eff}=0.6$ and we have assumed the ratio of EW breaking scales $f/v=3$.  The solid, dotted and dashed curves denote $\epsilon = 0,\,0.05$ and $0.08$ respectively, and the region with $\Delta N_\mathrm{eff}\le0.6$ lies below the corresponding curve. The red and green solid lines correspond to the width of the right handed neutrino in the model described in section~\ref{sec:neutrinomodel}, showing that this model produces a viable cosmology. 
}
   \label{fig:gammamregion}
   \end{figure}

In order to realize this scenario successfully, the following conditions 
must be satisfied:
 \begin{itemize} 
 \item{Most of the $N$ must have decayed before the temperature in the 
SM sector falls below an MeV, which is the temperature at which the SM 
neutrinos decouple from the thermal bath.}
 \item{Most of the decays of the $N$ must occur after the SM has 
decoupled from its twin counterpart. This is to ensure that the 
contribution to the energy density from the decays of $N$ is not shared 
equally between the two sectors.}
 \item{Finally, after the $N$ have decayed, the energy density in the SM 
sector must be at least an order of magnitude larger than in the twin sector,
in order to satisfy the bound on $N_\mathrm{eff}$.}
 \end{itemize}
 These conditions place constraints on the parameters of the theory which 
we now determine. In the regime where the energy density $\rho$ is 
dominated by the right-handed neutrinos, we have that
 \begin{equation}
3 H^2 M_{Pl}^2 = \rho = M_N n_N
 \end{equation}
 In this expression $n_N$ denotes the number density of the $N$, while 
$M_{Pl}$ represents the reduced Planck mass. Most of the $N$ decay 
close to the time when $H = \Gamma_{N}$. Assuming that $N$ decays 
predominantly to $A$ sector particles, $\epsilon \ll 1$, and working in the limit that all the $N$ decay instantaneously when 
$H = \Gamma_N$, we equate the total energy density in the $A$ sector 
immediately before and after these decays to obtain the relation,
 \begin{equation}
\rho_{A,R} = \frac{\pi^2}{30}{g}_{*A,R} T_{A,R}^4 
=  M_N n_N = 3 {\Gamma_N}^2 M_{Pl}^2\; .
\label{rhoAR}
 \end{equation}
 Here $T_{A,R}$ denotes the temperature in the SM sector immediately 
after the $N$ have decayed, and ${g}_{*A,R}$ represents the number of 
degrees in the visible sector at that temperature. In obtaining this 
expression we have neglected the small fraction of decays to the B 
sector, $\epsilon$. We can use Eq.~(\ref{rhoAR}) to obtain an expression 
for $T_{A,R}$ as a function of the width $\Gamma_N$ of the $N$,
 \begin{equation}
T_{A,R} = \left(
\frac{90 \Gamma_{N}^2 M_{Pl}^2}{{g}_{*A,R} \pi^2}
\right)^{\frac{1}{4}} \; .
 \label{TA,R}
 \end{equation}
 Requiring that $T_{A,R}$ lie above 1 MeV, we obtain a lower bound on 
$\Gamma_N$ represented by the lower horizontal line in figure~\ref{fig:gammamregion}.

 If the SM states are to be at a higher temperature than their twin 
counterparts, most of the $N$ must decay after the two sectors have 
decoupled. To satisfy the bound on $N_\mathrm{eff}$, the energy density 
in the $A$ sector must be at least an order of magnitude larger than in 
the $B$ sector. This implies that, at the time when the two sectors 
decouple, fewer than about 20\% of the $N$ must have decayed. It follows 
that at the decoupling temperature $T_D$,
 \begin{equation}
\frac{\Gamma_N}{H} \lesssim \frac{1}{5} \; .
 \label{GammaD}
 \end{equation}
 If the universe is still radiation dominated at this time, $H$ satisfies
 \begin{equation}
3 H^2 M_{Pl}^2 = \frac{\pi^2}{30} g_{*D} T_D^4
 \label{HubbleD}
 \end{equation}
Combining Eqs.~(\ref{GammaD}) and (\ref{HubbleD}), and taking $T_D$ to 
be 3 GeV, we obtain an upper bound on $\Gamma_N$ shown as the upper horizontal
 line in figure~\ref{fig:gammamregion}. Using Eq.~(\ref{TA,R}), this can be translated
into a bound on $T_{A,R}$,
 \begin{equation}
T_{A,R} \lesssim \frac{T_D}{2}
 \end{equation}
 The next step is to determine the ratio of the energy density in the $B$ 
sector, $\rho_B$, to the energy density in the $A$ sector, $\rho_A$, 
after the $N$ have decayed, and thereby obtain an expression for 
$N_\mathrm{eff}$. Our scenario assumes that the $N$ go out of the 
thermal bath when they are still relativistic. At the temperature $T_0$ 
when this happens, their number density is given by
 \begin{equation}
n_{N,0} = g_{*N} \frac{3\zeta(3)}{4\pi^2} T_0^3
 \end{equation}
 As the universe expands, the number density of the $N$ falls 
with the scale factor $a$ as
 \begin{equation}
n_N = n_{N,0} \frac{a_0^3}{a^3} \; .
 \end{equation}
 Here $a_0$ is the scale factor at temperature $T_0$. When the age of 
the universe approaches the lifetime of the $N$, so that $H = \Gamma_N$, 
we have
 \begin{equation}
3 \Gamma_N^2 M_{Pl}^2 = M_N n_{N,0}  \frac{a_0^3}{a_R^3} 
= M_N  g_{*,N} \frac{3\zeta(3)}{4\pi^2} T_0^3 \frac{a_0^3}{a_R^3} \; .
\label{a0aR}
 \end{equation}
 In this expression $a_R$ corresponds to the scale factor at the time 
when the $N$ decay, $H = \Gamma_N$. 

 By requiring that comoving entropy is conserved, we can obtain an 
expression for the total energy density in radiation at the time when 
the SM decouples from the twin sector,
 \begin{equation}
\rho_{D} =
\frac{\pi^2}{30}\left(\frac{g_{*,0}}{g_{*,D}}\right)^{1/3}
g_{*,0} T_0^4 \left(\frac{a_0}{a_D}\right)^4 
\; .
 \end{equation}
 Here $g_{*,0}$ and $g_{*,D}$ represent the total number of degrees of 
freedom in the bath at the temperatures $T_0$ and $T_D$, respectively, 
while $a_D$ denotes the scale factor at $T_D$. In this expression, we 
have neglected the small contribution to the energy density that arises 
from the decays of the $N$ prior to this time. As the temperature falls 
below $T_D$ this energy density is distributed between the SM and its 
twin counterpart, the relative fraction being determined by the number 
of degrees of freedom $g_{*A,D}$ and $g_{*B,D}$ in the SM and twin 
sectors at decoupling. The corresponding expressions for the energy 
densities in the SM and twin sectors immediately after decoupling are 
given by
 \begin{equation}
 \rho_{A,D} = \frac{g_{*A,D}}{g_{*,D}} \rho_D \qquad \mbox{and} \qquad
 \rho_{B,D} = \frac{g_{*B,D}}{g_{*,D}} \rho_D \,.
 \end{equation}
 As the system evolves and cools, species continue to go out of both the 
$A$ and $B$ baths, resulting in an increase in the comoving energy 
density of the corresponding sectors. Eventually the $N$ decay, giving an 
especially large contribution to the energy density of the SM sector.
 
The total energy density in the $A$ sector immediately after the $N$ 
have decayed, in the instantaneous decay approximation, is given by 
Eq.~(\ref{rhoAR}). The corresponding energy density in the $B$ sector 
can be approximated as
 \begin{equation}
\rho_{B,R} = g_{*B,R} T_{B,R}^4 = 
3 \epsilon {\Gamma_N}^2 M_{Pl}^2 + 
\left(\frac{g_{*B,D}}{g_{*B,R}}\right)^{1/3} \rho_{B,D} 
\left(\frac{a_D}{a_R}\right)^4\; .
\label{rhoBR}
 \end{equation}
 Here $T_{B,R}$ denotes the temperature in the $B$ sector immediately 
after the $N$ have decayed and $g_{*B,R}$ the number of degrees of 
freedom in the $B$ sector at that temperature. In this expression, the 
first term on the right hand side represents the contribution to the 
energy density arising from the decays of the $N$. The second term is 
independent of the $N$ and is instead associated with the primeval 
energy density in the $B$ sector. Since $\epsilon$ is small this term 
cannot, in general, be neglected. Taking the ratio of Eqs.~(\ref{rhoAR}) 
and (\ref{rhoBR}), and using Eq.~(\ref{a0aR}) to eliminate the ratio of 
scale factors in favor of the width $\Gamma_N$ we obtain,
  \begin{equation}
\frac{\rho_B}{\rho_A} = \epsilon +  R_N \; ,
 \end{equation}
 where $R_N$ is given by
 \begin{equation}
R_N = \frac{\pi^2}{90} \left(\frac{4\pi^2}{\zeta(3)}\right)^{4/3}
\left(\frac{g_{*B,D}}{g_{*N}}\right)^{4/3}
\left(\frac{g_{*,0}}{g_{*,D}}\right)^{4/3}
\left(\frac{\Gamma_N^2 M_{Pl}^2}{g_{*B,R} M_N^4}\right)^{1/3} \; .
\label{eq:RN}
 \end{equation}
 We see that for our mechanism to be effective, both $\epsilon$ and $R_N$ 
are required to be small. 

Between the time of $N$ decay and late times, \ie\ BBN and CMB, the $A$ 
and $B$ sectors may pass through additional mass thresholds, each of 
which results in an increase of $(g_*^{(i)}/g_*^{(f)})^{1/3}$ in the 
temperature and energy density of the corresponding sector.  In order to 
compare to the limit on $\Delta N_\mathrm{eff}$ we compare the energy density 
in the $B$ sector to that in SM neutrinos at the temperature of neutrino 
decoupling, Eq.~(\ref{eq:neffatMeV}). This leads to the expression
 \be
 \label{eq:Neff}
\Delta N_{\rm eff} \approx 7.4 
\left(\frac{g_{*B,R}}{g_{*A,R}}\right)^{1/3}
\left(\epsilon + R_N\right)~.
 \ee
 In figure~2 we have plotted $\Delta N_\mathrm{eff}$ as a function of $M_N$ and 
$\Gamma_N$ for different values of $\epsilon$. We see that provided 
$\epsilon \lesssim 1/10$, there is a broad range of parameters where the 
constraints from cosmology on $\Delta N_\mathrm{eff}$ can be satisfied. 
Moreover, while the dark radiation in the cooler twin sector satisfies 
current bounds on $\Delta N_\mathrm{eff}$, it may lie within reach of future 
measurements of $\Delta N_\mathrm{eff}$ such as CMB Stage-IV experiments 
\cite{CMB-S4:2016}.
   
The neutrinos in the twin sector will also contribute to the total mass 
density in neutrinos as measured by cosmology. We now seek to determine 
the magnitude of this effect. As we will see, this effect can be large, 
and constitutes a striking signal of this scenario. The first step is to 
determine the number density of neutrinos in the twin sector. Shortly 
prior to the decoupling of the SM and twin neutrinos from their 
respective baths, the degrees of freedom in the two sectors are 
identical, and consist of the electron, the photon and the three 
neutrinos. Then the ratio of neutrino number densities in the two 
sectors is given by
 \begin{equation}
\frac{n_{\nu,B}}{n_{\nu,A}} = \frac{T_B^3}{T_A^3} \; .
 \end{equation}
 Now, comoving entropy conservation implies that
 \begin{equation}
\frac{{g}_{*B,R} T_{B,R}^3}{{g}_{*A,R} T_{A,R}^3} =
\frac{T_B^3}{T_A^3} \; .
 \end{equation}
 We therefore find that
 \begin{equation}
\frac{n_{\nu,B}}{n_{\nu,A}} =
\frac{{g}_{*B,R} T_{B,R}^3}{{g}_{*A,R} T_{A,R}^3} =
\left(\frac{{g}_{*B,R}}{{g}_{*A,R}}\right)^{1/4} 
\left(\epsilon + R_N \right)^{3/4}~.
 \end{equation} 
 The comoving number density of neutrinos does not change during or 
after their decoupling from the thermal bath. Therefore this equation 
continues to remain true at late times, and can be used to determine the 
number density of twin neutrinos.

It follows from this that the ratio of the mass densities in neutrinos 
satisfies 
 \begin{equation}
\frac{m_{\nu,B} n_{\nu,B}}{m_{\nu,A} n_{\nu,A}}
\approx \frac{m_{\nu,B}}{m_{\nu,A}}\left(\frac{\Delta N_{\mathrm{eff}}}{7.4}\right)^{3/4}~.
 \end{equation} 
 In the absence of additional $\mathbb{Z}_2$ breaking in the neutrino sector, 
twin neutrinos are expected to be heavier than their SM counterparts. 
The reason is that neutrino masses arise as an electroweak symmetry 
breaking effect, and the vev of the Twin Higgs is larger than that of 
the SM Higgs. If neutrinos are Dirac, we expect that 
$m_{\nu,B}/m_{\nu,A} \approx f/v$. If instead neutrinos are Majorana, we 
expect that $m_{\nu,B}/m_{\nu,A} \approx f^2/v^2$, in the absence of any 
$\mathbb{Z}_2$ breaking in the neutrino sector. We see that if neutrinos are 
Majorana, for $\epsilon = 1/10$ and $v/f < 1/3$, the total mass density 
in neutrinos is larger than in the SM. It is important to keep in mind 
that the bounds in the literature on the sum of neutrino masses are not 
directly applicable, since the twin neutrinos are at a lower temperature 
than the SM neutrinos. Nevertheless, this constitutes a striking signal 
of this scenario.

\section{Neutrino Masses and Cosmology
} 
\label{sec:neutrinomodel} 

 In this section we extend the MTH framework to include neutrino masses 
by incorporating into the theory a Type-I seesaw. 
We show that if the 
mass scale of the right-handed neutrinos is of order a GeV, this 
construction offers a simple resolution to the cosmological problems 
associated with this class of models along the lines discussed in 
Section~\ref{sec:cosmology}, and leads to interesting predictions for upcoming 
experiments. 
We begin in~\S\ref{sec:onefam} with a toy model in which there is only one family of neutrinos and show that the allowed parameter space is accessible. We later show in~\S\ref{sec:threefam} that this is also the case in two classes of models with three families. 
Since, in detail, the phenomenology depends on whether or 
not the neutrino sector respects the $\mathbb{Z}_2$ twin symmetry, we briefly consider 
the possibility of $\mathbb{Z}_2$ breaking in~\S\ref{sec:Z2breaking}.

\subsection{$\mathbb{Z}_2$ Symmetric Neutrino Sector with One Family}\label{sec:onefam}
 
To illustrate the mechanism, we consider first the case of just one 
family of SM and twin neutrinos. The relevant terms in the Lagrangian 
take the schematic form,
 \begin{equation}
\mathcal{L} \supset - y \left( L_A H_A N_A + L_B H_B N_B \right) 
- \frac{1}{2} M_N \left( N_A^2 + N_B^2 \right) -  M_{AB} N_A N_B + 
{\rm h.c.} \; \label{eq:Nmass}
 \end{equation}
 Here the subscripts $A$ and $B$ denote the SM fields and their twin 
counterparts respectively. The discrete $\mathbb{Z}_2$ symmetry enforces the 
equality of the mass and interaction terms in the SM and twin sectors. 
We have included a mass parameter $M_{AB}$ that mixes the right-handed 
neutrinos in the two sectors. In what follows, we assume a hierarchy in 
the parameters $M_N \gg M_{AB} \gg y \langle H \rangle $. Then, because 
of the $\mathbb{Z}_2$ symmetry, the mass eigenstates in the right-handed neutrino 
sector are given by,
 \begin{eqnarray}
N_+ &=& \frac{1}{\sqrt{2}} \left( N_A + N_B \right) \nonumber \\
N_- &=& \frac{1}{\sqrt{2}} \left( N_A - N_B \right) \; . \label{eq:Neigen}
 \end{eqnarray}  
 The corresponding mass eigenvalues are given by $M_{\pm} = M_N \pm M_{AB}$.

Integrating out the right-handed neutrinos, we obtain expressions for 
the neutrino masses, 
 \begin{eqnarray}
m_{\nu, A} &=& \frac{y^2 \langle H_A \rangle^2}{M_N} 
\left\{ 1 + \mathcal{O}\left( \frac{M_{AB}}{M_N} \right) \right\}
\nonumber \\
m_{\nu, B} &=& \frac{y^2 \langle H_B \rangle^2}{M_N}
\left\{ 1 + \mathcal{O}\left( \frac{M_{AB}}{M_N} \right) \right\}
\; .
 \end{eqnarray}
 We see that, even though the right-handed neutrino mass eigenstates 
consist of an equal mix of visible and mirror states, the final result 
for the neutrino mass in the SM sector is exactly as expected from the 
familiar Type-I seesaw, up to small corrections that arise as a 
consequence of mixing between the $A$ and $B$ sectors. The neutrino mass 
eigenstates in each sector also contain a small 
$\mathcal{O}\left({M_{AB}}/{M_N} \right)$ admixture of neutrinos from 
the other sector. Provided this mixing is $\lsim10^{-3}$ the bounds 
arising from the oscillations of active neutrinos into sterile twin 
states in the early universe~\cite{Hannestad:2012ky} can be satisfied.

We focus on a region of parameter space in which the right-handed 
neutrino mass $M_N$ is of order a GeV, while $M_{AB}$ is of order 
an MeV. Then, in order to reproduce neutrino masses in the range from 
$10^{-3} - 10^{-1}$ eV, $y$ is expected to be of order $10^{-7} - 
10^{-8}$.

The right-handed neutrinos $N_+$ and $N_-$ can decay into visible sector 
fermions through the weak interactions.\footnote{They may also decay through the Higgses $h_A$ and $h_B$, but these decays are suppressed by small Yukawa couplings and are numerically subdominant to weak decays.}  Both charged and neutral 
currents contribute. We can estimate the decay width as
 \begin{equation}
 \label{eq:width}
\Gamma_{N \rightarrow A} \approx C_{A} \frac{G_F^2}{192 \pi^3} 
\left(\frac{m_{\nu, A}}{M_N}\right) M_N^5 
\; .
 \end{equation}
Here $C_{A}$ involves a sum of order-one numbers that account for the multiplicity of final states, and we have neglected the masses of the final state particles. 
The right-handed 
neutrinos can also decay into hidden sector fermions through the weak 
interactions in the twin sector. However, these decay modes are 
suppressed because the weak gauge bosons in the twin sector are heavier 
by a factor $(f^2/v^2)$ than the corresponding particles in the SM. The 
corresponding decay width can be estimated as
 \begin{equation}
\Gamma_{N \rightarrow B} \approx C_{B} \frac{G_F^2}{192 \pi^3}
\left(\frac{m_{\nu, B}}{M_N}\right) \left(\frac{v}{f}\right)^4
M_N^5 
\; .
 \end{equation}
 The parameter $C_{B}$ again involves a sum over order-one numbers. In 
the limit that the masses of the final state particles are neglected, 
and the same number of decay channels are open in the two sectors, we 
have that $C_{A} = C_{B}$. We see that decays into twin states are 
suppressed because the $W$ and $Z$ gauge bosons in the $B$ sector are 
heavier by a factor of $f^2/v^2$ than in the $A$ sector. Although the 
neutrino mass in the $B$ sector is heavier by a factor $f^2/v^2$, 
leading to an enhancement, this is not sufficient to compensate for the 
$(v^2/f^2)^2$ suppression that arises from the hierarchy in gauge boson 
masses. The fraction of right-handed neutrino decays into hidden sector 
states can be estimated as
 \begin{equation}
\epsilon =  \frac{\Gamma_{N \rightarrow B}}{\Gamma_N} \approx
\frac{v^2}{f^2}~.
 \end{equation} 
 It follows that for $v/f$ of order $1/5$, the width into twin states 
can be as small as a few percent. We see that even in the absence of any 
additional breaking of the $\mathbb{Z}_2$ symmetry in the neutrino sector, it is 
strightforward to obtain small values of $\epsilon$.

We now show that this simple mechanism for neutrino masses can indeed 
lead to a viable cosmology using the mechanism described in the previous section. For this to work we must show - (a) that the right handed neutrinos decouple when they are relativistic, and (b) that the lifetime of right-handed neutrinos fit within the allowed region of figure~\ref{fig:gammamregion}.

At temperatures below the weak scale the 
right-handed neutrinos $N_+$ and $N_-$ are kept in chemical equilibrium 
by the weak interactions through processes such as $N + e^- \rightarrow 
\nu + e^-$. 
These process will eventually freeze out at the temperature $T_0$. To estimate this freeze out temperature we can scale up the freeze out temperature of regular SM neutrinos in standard cosmology as follows. First we note that as long as $N$ is relativistic  
\begin{equation}
\frac{ \sigma(N + e^- \rightarrow \nu + e^-)}{\sigma(\nu + e^- \rightarrow \nu + e^-)} \sim
\frac{m_\nu}{M_N}\,.
\end{equation}
Following the standard procedure of equating the interaction and the expansion rate, we find that the decoupling temperature $T_0$ scales as
\begin{equation}
T_0 \sim \left(\frac{M_N}{m_\nu}\right)^{1/3} T_{\nu,\mathrm{SM}}
\label{eq:Ndecouplingtemp}
\end{equation}
where $T_{\nu,\mathrm{SM}}\sim1$ MeV is the temperature of neutrino decoupling in the SM. 
These processes freeze out at a temperature of order 10 GeV, 
when the $N$ are still relativistic.  Requiring that the right handed neutrinos decouple when they are relativistic, $T_0>M_N$, we find a constraint
\begin{equation}
M_N< 10\mbox{ GeV} \left(\frac{0.01\mbox{ eV}}{m_\nu}\right)^{1/2} \,,
\end{equation}
which is easy to satisfy and will, in fact, be less stringent than the requirement on the right handed neutrino lifetime.
 
Assuming this constraint is satisfied, as the universe continues to  expand, $N_+$ and $N_-$ become nonrelativistic, and eventually come to  dominate the energy density of the universe. Finally, when $\Gamma_N 
\approx H$, the right-handed neutrinos decay, contributing to the 
entropy of the SM and twin sectors. 
Since $\epsilon$ is small, the SM is 
heated up more than the twin sector, allowing the cosmological bounds to 
be satisfied. 
For example, setting $\epsilon$ to zero, and combining the equations~(\ref{eq:RN}) and~(\ref{eq:Neff}) for $\Delta N_\mathrm{eff}$ with equation~(\ref{eq:width}) for the width, requiring $\Delta N_\mathrm{eff}<0.6$ implies
\begin{equation}
M_N< 1\mbox{ GeV} \left(\frac{0.01\mbox{ eV}}{m_\nu}\right)^{1/2} \,,
\end{equation}
We thus find that in our neutrino model the requirement of the right handed neutrinos to dominate the energy density of the Universe to sufficient degree before they decay is more stringent that that for relativistic decoupling. In fact, we find that for a sufficiently light left handed neutrino all three constraints on the width of the right handed neutrino can be satisfied, as in shown in figure~\ref{fig:gammamregion}, where the red and green lines which represent the mass-width relation for neutrino masses of $10^{-4}$ and $10^{-6}$ eV traverse the allowed region.

\subsection{$\mathbb{Z}_2$ Violating Neutrino Sector}\label{sec:Z2breaking}

 The discussion above shows that in the case when the neutrino sector 
respects the $\mathbb{Z}_2$ symmetry, the branching fraction of $N$ decays into 
the twin sector is given by $\epsilon = v^2/f^2$, ignoring the effects of different final state particle masses in the two sectors. If, however, the 
neutrino sector explicitly violates the $\mathbb{Z}_2$ symmetry, much smaller 
values of $\epsilon$ can be accommodated. Perhaps the most 
straightforward way to suppress $\epsilon$ is to set the Yukawa 
couplings of the neutrinos in the $B$ sector to zero. The result of this 
would be to turn off the decays of right-handed neutrinos into $B$ 
sector states, effectively setting $\epsilon=0$. Although this 
constitutes a hard breaking of the $\mathbb{Z}_2$ symmetry, since the Yukawa 
couplings in Eq.~(\ref{eq:Nmass}) are so small, this has only a 
negligible effect on the Higgs potential.

One can also introduce soft $\mathbb{Z}_2$ breaking in order to suppress 
$\epsilon$. For example, in Eq.~(\ref{eq:Nmass}) we assumed that the 
mass of $N$ respects $\mathbb{Z}_2$ exactly. We can instead relax this 
requirement, giving the two right handed neutrinos different 
masses
 \begin{equation}
\mathcal{L} \supset 
- \frac{1}{2} M_{N_A} N_A^2 - \frac{1}{2} M_{N_B} N_B^2  -  M_{AB} N_A N_B + 
{\rm h.c.} \; \label{eq:Nmass}
 \end{equation}
This $\mathbb{Z}_2$ 
breaking effect could either be introduced as a soft breaking ``by 
hand'' or as a result of the existing $\mathbb{Z}_2$ breaking of $v<f$. For 
example, the $\mathbb{Z}_2$ symmetric dimension 5 operator
\begin{equation}
\mathcal{L} \supset \frac{1}{\Lambda}\left( |H_A|^2 N_A^2+ |H_B|^2 N_B^2\right)
\end{equation}
leads to a splitting of $(f^2-v^2)/\Lambda$, once $H_{A,B}$ are set to their vevs.
Now, the mass eigenstates for right handed neutrinos will no longer be the maximal mixture of Eq.~(\ref{eq:Neigen}), but would rather involve a mixing angle $\theta$:
\begin{eqnarray}
N_1&=&\cos\theta\, N_A + \sin\theta\, N_B \nonumber \\
N_2&=&\cos\theta\, N_B - \sin\theta\, N_A ~,
\end{eqnarray}
where $\tan 2\theta=2M_{AB}/(M_B^2-M_A^2)$. If $M_B>M_A$, $N_1$ will be lighter than $N_2$. The result is that the lightest right handed neutrino is mostly part of the $A$ sector and its branching ratio into the $B$ sector $\epsilon$ will be further suppressed by $\sin^2\theta$. Of course, in this case $N_2$ would be mostly in the $B$ sector and would have an enhanced branching to $B$ states, however because it is heavier it can be significantly shorter-lived and would thus have a lesser impact on cosmology. We leave the detailed analysis of such a framework for future study and in this work will simply keep in mind that $\epsilon$ is a free parameter which is motivated to be of order $v^2/f^2$ but could also be smaller.

\section{The $\nu$MTH Model}\label{sec:threefam}

So far we have dealt with a toy model for the neutrino sector, one which has just one flavor. In this simple case the width of the right handed neutrino is proportional to the mass of the light active neutrino as in equation~(\ref{eq:width}). We found that there is a viable cosmology if the mass of the light left-handed neutrino is sufficiently small. In generalizing our model to the three neutrino case one may expect a gain, since now three right handed neutrinos will be decaying preferentially into the visible sector, leading to a larger asymmetry in energy densities. It remains to be seen whether one can maintain and improve upon the  success of the one flavor model  while requiring full agreement with neutrino oscillation data. In particular, it is interesting to keep track of which light neutrino mass will be suppressing the width of the various $N$'s. Is it the lightest neutrino, the heaviest, or a linear combination? We will find that the answer depends on the flavor structure of the neutrino sector, and that all of these are a possibility.

We now discuss the $\nu$MTH which extends the framework to three flavors, starting with a general treatment and then giving two examples which produce different behaviors.
The relevant 
terms in the 3-flavor Lagrangian take the form,
 \begin{equation}
\mathcal{L} \supset - y_{ij} \left( L_A^i H_A N_A^j + L_B^i H_B N_B^j \right) 
- \frac{1}{2} (M_N)_{ij} \left( N_A^i N_A^j + N_B^i N_B^j \right) -  (M_{AB})_{ij} N_A^i N_B^j + 
{\rm h.c.} \; 
 \end{equation}
 Here the subscripts $i,j = 1,2,3$ denote the three generations in the 
gauge eigenbasis. Although at this stage we could go directly to the 
mass eigenbasis by diagonalizing the full neutrino mass matrix, it is 
more illuminating to proceed in a series of steps. Once again assuming 
the hierarchy $M_N \gg M_{AB} \gg y \langle H \rangle$, and neglecting 
the small corrections from electroweak symmetry breaking, the mass 
matrix for the right-handed neutrinos $(N_A^i, N_B^j)$ takes the form
 \begin{eqnarray}
 \mathcal{M} = \left( \begin{array}{cc} M_N & M_{AB} \\ M_{AB}^T &  M_N \end{array} \right) \, .
 \end{eqnarray}
 This mass matrix can be block-diagonalized by means of an orthogonal rotation
bringing it into the form,
 \begin{eqnarray}
 \left(\begin{array}{cc}   M_N + \frac{1}{2} (M_{AB} + M_{AB}^T) & 0 \\ 0 &  M_N - \frac{1}{2} (M_{AB} + M_{AB}^T) \end{array} \right) = O^T \mathcal{M} O \; \; .
 \end{eqnarray}
 This rotation takes the right-handed neutrino states from the gauge 
eigenbasis to the approximate sterile neutrino mass eigenstates 
$N_{\pm}^i = \frac{1}{\sqrt{2}} (N_{A}^i \pm N_{B}^i)$, the natural 
generalization of the one-flavor case. Electroweak symmetry breaking 
then induces small off-diagonal terms that mix the left-handed and 
right-handed neutrinos. At leading order, this gives the active neutrino 
mass matrices
\begin{eqnarray}
(m_{\nu, A})_{ij} &=& -  \frac{1}{2} (\theta_A)_{ik} (M_N)_{kl} (\theta_A)_{lj}^T
\left\{ 1 + \mathcal{O}\left( \frac{M_{AB}}{M_N} \right) \right\}
\nonumber \\
(m_{\nu, B})_{ij} &=&  -  \frac{1}{2} (\theta_B)_{ik} (M_N)_{kl} (\theta_B)_{lj}^T
\left\{ 1 + \mathcal{O}\left( \frac{M_{AB}}{M_N} \right) \right\}
\; .
 \end{eqnarray}
where e.g. $(\theta_A)_{ij} = y_{ik} (M_N^{-1})_{kj} \langle H_A \rangle$ and the right-handed neutrino mass matrices are only shifted at $\mathcal{O}(m_\nu / M_N)$. The matrices $m_{\nu, A}, m_{\nu,B}$ and $M_N$ can then be fully diagonalized by matrices $U_{\nu,A}, U_{\nu,B},$ and $U_N$, respectively, where $U_{\nu,A}$ can be identified with the PMNS matrix in a basis where the charged lepton mass matrix is already diagonal. In this expansion, the mixing angles between sterile neutrinos and active neutrino gauge eigenstates in the $A$ and $B$ sectors are given by 
\begin{eqnarray} \label{eq:angles}
(\Theta_A)_{ij} &\simeq& (\theta_A)_{ik} (U_N)_{kj} =y_{ik} (M_N^{-1})_{kl} (U_N)_{lj} \langle H_A \rangle \\
(\Theta_B)_{ij} &\simeq& (\theta_B)_{ik} (U_N)_{kj} = y_{ik} (M_N^{-1})_{kl} (U_N)_{lj} \langle H_B \rangle
\end{eqnarray}
with corrections of order $\mathcal{O}(M_{AB}/M_N)$. Note that, as in the single-family case, these mixing angles are naturally $\mathcal{O}(\sqrt{m_\nu/M_N} )$.
These mixing angles, squared and summed over the light neutrino index, enter the width of the right handed neutrinos. 
We reserve a detailed discussion of the decay widths for Appendix \ref{app:decays}.
We now consider two possible limits for the right handed neutrino sector - one in which right handed neutrinos are universal and thus aligned with the light neutrinos, and another in which they are anarchic.

\subsection{Universal Right Handed Neutrinos}\label{sec:universal}

A particularly simple limit of the three-family setup described above is one in which both $M$ and $M_{AB}$ are universal and $y$ is real,
\begin{equation}
M_N\propto \mathds{1} \qquad\mbox{and}\qquad M_{AB}\propto \mathds{1} \qquad\mbox{and}\qquad y_{ij}\in \mathbb{R}~.
\label{eq:universallimit}
\end{equation}
This limit is protected by a CP symmetry in the lepton sector, and an SO(3) symmetry which is only broken by the spurion $y_{ij}$.  We will still maintain the hierarchy $M_N\gg M_{AB}\gg\, y \langle H \rangle$ for simplicity.   In this case the mass basis for right and left handed neutrinos are aligned and the decay of each right handed neutrino is suppressed by a definite left handed neutrino mass. In the language of the previous subsection, in this limit the $U_N$ matrices are the identity and the $\theta$ matrices are simply diagonal matrices times the PMNS matrix (with $\delta_{CP}=0$),
\begin{equation}
\theta_{A} = \mathrm{diag}\left[\sqrt{\frac{m^{i}_{\nu,A}}{M_N} }\,\right] U_{\nu,A}\,.
\end{equation}
In the limit of massless decay products the PMNS matrix will drop out of the decay width once the amplitude is squared due to unitarity. 
As a result, the decay rates of the right handed neutrinos takes a form which is very similar to equation~(\ref{eq:width})
\begin{equation}
\label{eq:width3fam}
\Gamma_{N_i \rightarrow A} \approx C_{A} \frac{G_F^2}{192 \pi^3} 
\left(\frac{m^i_{\nu, A}}{M_N}\right) M_N^5 
\; .
 \end{equation}
where the decay of each $N_i$ is suppressed by its own light neutrino mass $m^i_{\nu, A}$. We see that in this universal case, as the lightest SM neutrino is taken to approach zero mass, one right handed neutrino would become arbitrarily narrow. In particular, we find that we are able to raise the RH neutrino mass $M_N$ while keeping the width of one RH neutrino fixed, thus moving to the right in the $M_N$-$\Gamma_N$ plane for this state, staying within the allowed region in figure~\ref{fig:gammamregion}. When we do so, the width of the other two right hand neutrinos cannot be held fixed because the masses of the corresponding left handed neutrinos cannot be taken to zero without coming in conflict with the measured mass difference measured in oscillation. 

\begin{figure}[t] 
   \centering
   \includegraphics[width=0.475\columnwidth]{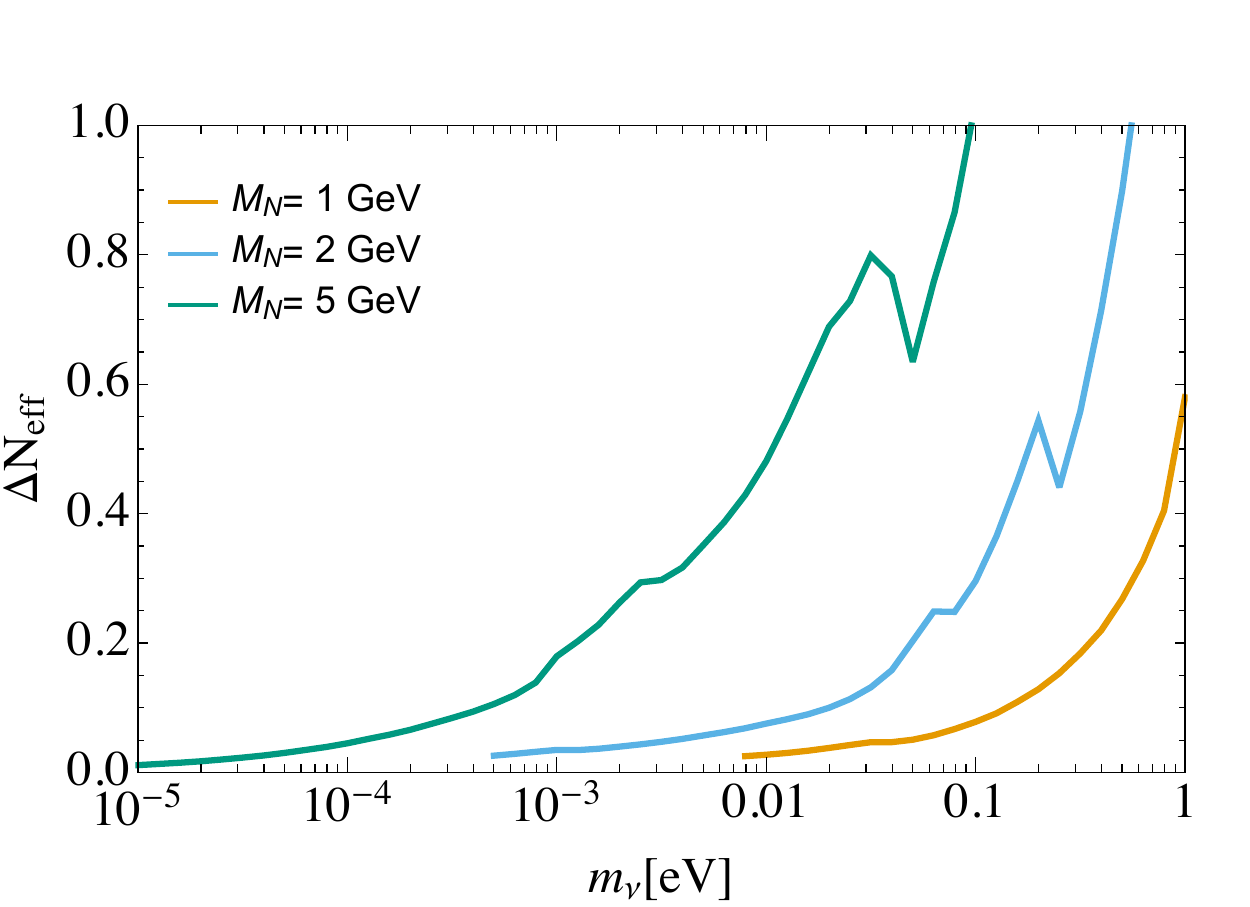} 
   \includegraphics[width=0.475\columnwidth]{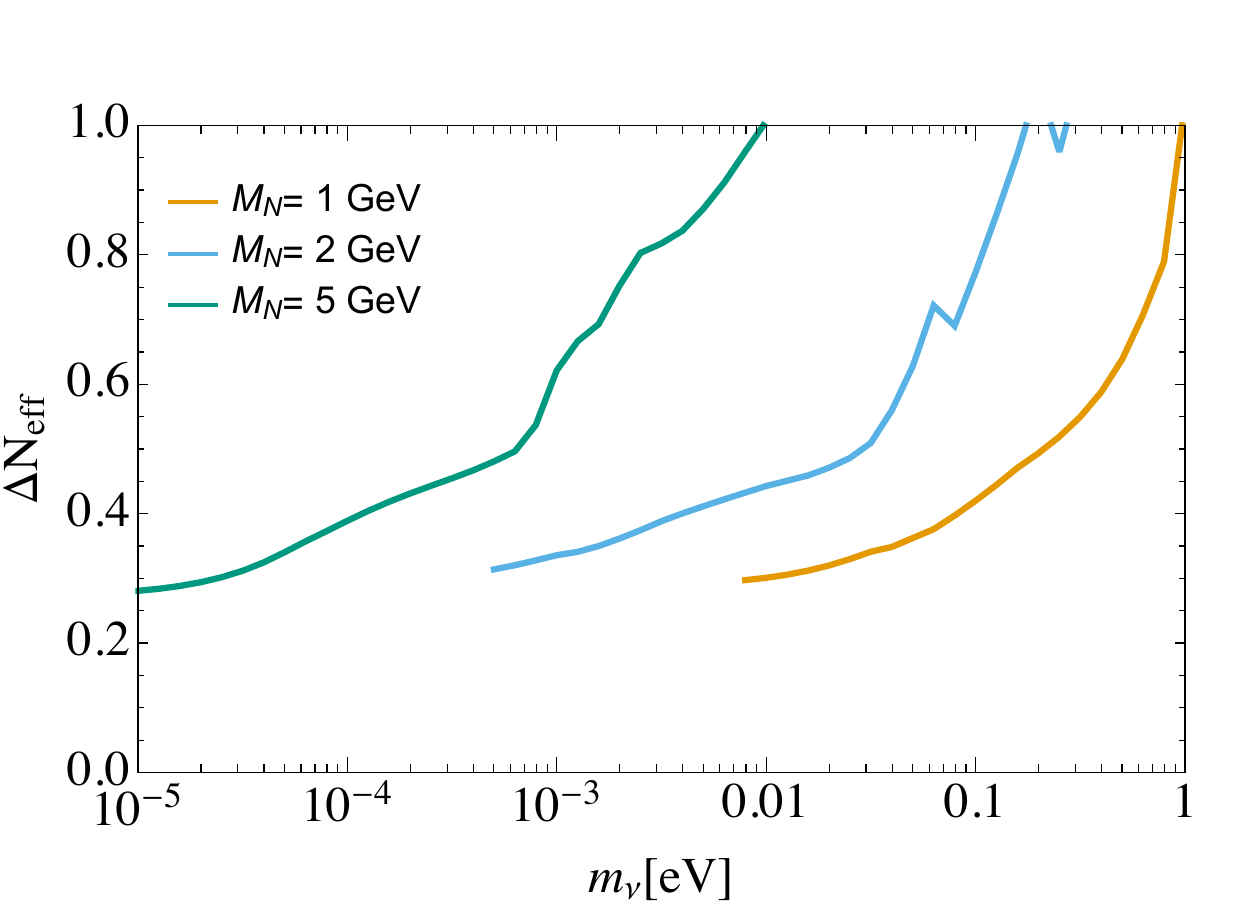} 
  \caption{$\Delta N_\mathrm{eff}$ as a function of the mass of the lightest left handed SM neutrino in the universal right handed neutrino limit. Curves are shown for right handed neutrinos at 1, 2, and 5~GeV. Left plot is for $\epsilon=0$ while the right plot is for $\epsilon=0.05$.} 
   \label{fig:universalplot}
   \end{figure}

In the limit that two right handed neutrinos decay early and do not affect the cosmology, we thus find that the limit of (\ref{eq:universallimit}) effectively reproduces the results of the single family result of section~\ref{sec:onefam} and figure~\ref{fig:gammamregion}. However, even in this case it is possible for all right handed neutrinos to contribute to the temperature difference between the A and the B sectors. In figure~\ref{fig:universalplot} we show the contribution to $\Delta N_\mathrm{eff}$ from twin states as a function of the lightest SM neutrino mass. In the figure $\Delta N_\mathrm{eff}$ was estimated numerically, going beyond the instantaneous decay approximation and accounting for the decays of the various $N$'s in different times, as described in Appendix~\ref{app:Numerical}. We assume that the $B$ sector particles are three times heavier than their twins in the SM, \ie\ $f/v=3$.  We consider two choices for the branching ratio into the twin sector: $\epsilon=0.05$, which is slightly smaller than expected from $\mathbb{Z}_2$ alone, and the limiting case of $\epsilon=0$ where the only contribution to $\Delta N_{\mathrm{eff}}$ comes from the primeval energy density, $R_N$. We find that there is ample regions of parameter space where $\Delta N_\mathrm{eff}$ is acceptably small.

\subsection{Anarchic Right Handed Neutrinos}
\label{sec:Guidology}

In the previous subsection we assumed a flavor universal right handed neutrino sector which allowed for an arbitrarily narrow species of right handed neutrino, in the limit of a very light left-handed neutrino mass. This alignment could be guaranteed using a flavor and CP symmetry. One would expect that if the right handed neutrino masses were flavor anarchic and CP violating this one-to-one correspondence of right and left handed neutrino mass eigenstates would be spoiled. In this case the equivalent of equation~(\ref{eq:width3fam}) will include a linear combination of all three light neutrino masses which cannot be arbitrarily small and as a result the right handed neutrinos will all have a minimal generic width. Does this imply then that our mechanism can only work for the restrictive aligned models? 
We now show that this is \emph{not} the case and even an anarchic neutrino sector can lead to a viable Twin Higgs cosmology.
As an example, we present here a specific example of a seesaw model of neutrino masses and mixings, based around ``pseudo $\mu\tau$-anarchy''~\cite{Altarelli:2002sg,Altarelli:2012ia}, and demonstrate that it can simultaneously give the correct SM neutrino parameters and realize lifetimes for the RH neutrinos that are sufficiently long to satisfy the constraints discussed in section~\ref{sec:cosmology}. 

To generate a partially anarchical texture for the neutrino masses the 
model consists of two flavons of opposite $U(1)_{FN}$ charge, that 
acquire an equal vev.  In the left-handed lepton sector the first 
generation is taken to have $U(1)$ charge 2, while the second and third 
generation have no charge.  In the right-handed neutrino sector the 
first and second generation have charges $\pm 1$ respectively, while the 
third generation is neutral\footnote{Note that we neglect the 
tiny effect of 
the $M_{AB}$ terms on the masses and mixings of the light neutrinos.  
In the notation of section~\ref{sec:cosmology} 
$N=N_\pm$.}.  This results in textures of the form
 \be
m_D = \overline{m}_D \begin{pmatrix}
\lambda^3 & \lambda  & \lambda^2 \\
\lambda     & \lambda  & 1 \\
\lambda    & \lambda    & 1
\end{pmatrix}
,
\quad
M_N = \overline{M}_N \begin{pmatrix}
\lambda^2 & 1                & \lambda \\
1               & \lambda^2  & \lambda \\
\lambda    & \lambda      & 1
\end{pmatrix}~.
\ee
The Dirac mass term for the $B$ sector is $f/v$ larger than in the $A$ sector.
We then generate random matrices with each entry, $m_{ij}$, picked uniformly from $[0.5,2]\times t_{ij}$ where $t_{ij}$ is the corresponding texture entry.  In addition each entry acquires a random phase from $0$ to $2\pi$, and $M_N$ is symmetrised.  Following \cite{Altarelli:2012ia} we take $\lambda\approx 0.35$.  After diagonalising the full neutrino mass matrix, we require that the resulting neutrino mixing parameters and mass splittings for the SM neutrinos are within $3\sigma$ of the best fit values presented in the PDG \cite{Agashe:2014kda}:
\bea
2.23\times 10^{-3}\, \eV^2 \le & \Delta m^2_{atm}& \le 2.61 \times 10^{-3}\, \eV^2 \nonumber \\
6.99 \times 10^{-5}\, \eV^2 \le & \Delta m^2_{sol}& \le 8.18 \times 10^{-5}\, \eV^2 \nonumber\\
0.259 \le & \sin^2\theta_{12}& \le 0.359\nonumber \\
0.374 \le &\sin^2\theta_{23}& \le 0.628 \nonumber\\
0.0176 \le & \sin^2\theta_{13}& \le 0.0295~.
\label{eq:SMneutrinovalues}
\eea
We only consider the case of normal ordering of the neutrino masses where $\Delta m^2_{atm} \equiv m_3^2-(m_1^2+m_2^2)/2$ and $\Delta m^2_{sol}\equiv m^2_2-m_1^2$ are both positive.  The active neutrinos have masses that scale as $m_\nu\sim m_D^2/M_N$ so that any particular realisation of the textures that satisfies (\ref{eq:SMneutrinovalues}) is actually a one-parameter family of solutions, where $m_D\rightarrow r^{1/2}m_D$ and $M_N\rightarrow r M_N$.

\begin{figure}[t] 
   \centering
   \includegraphics[width=0.475\columnwidth]{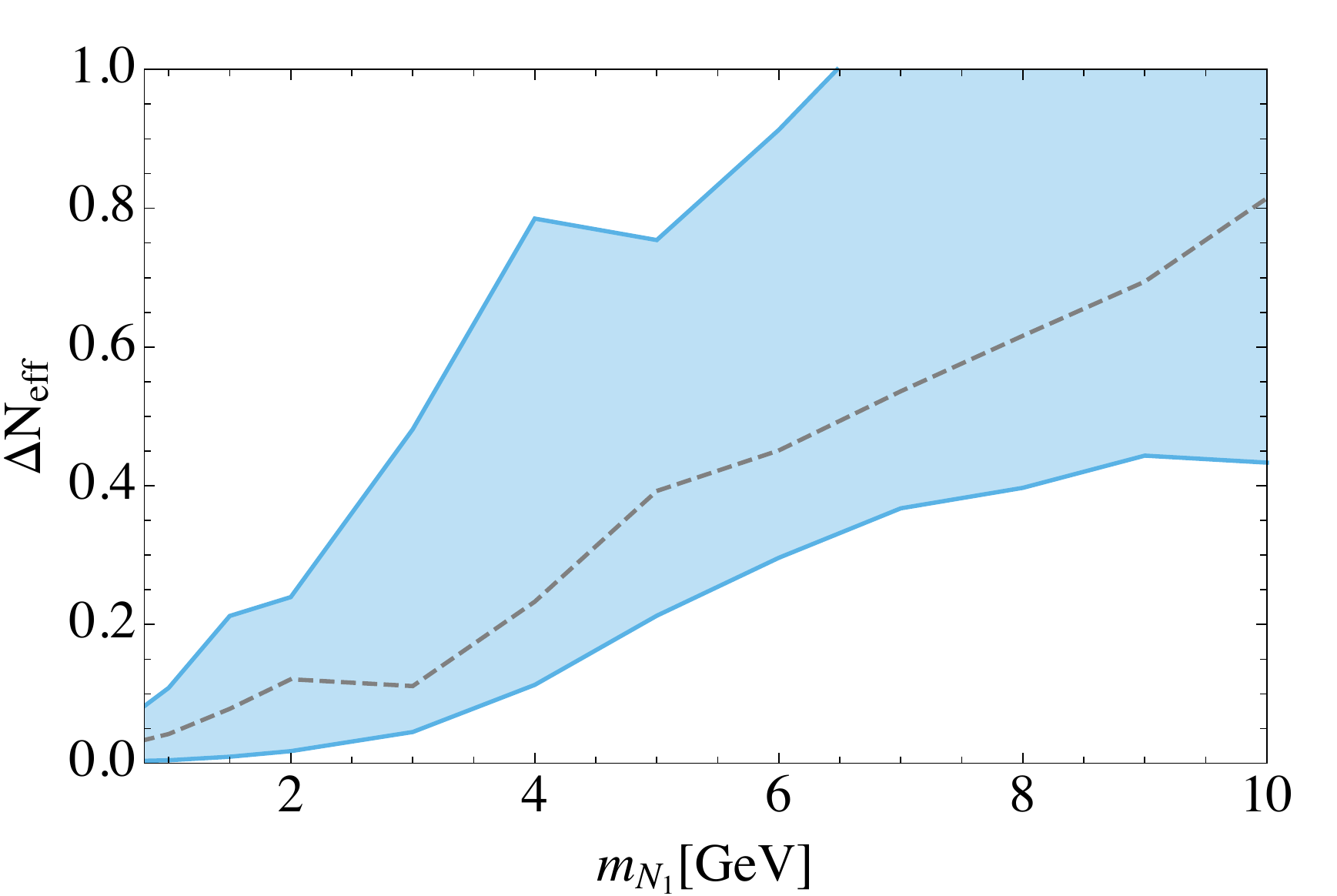} 
   \includegraphics[width=0.475\columnwidth]{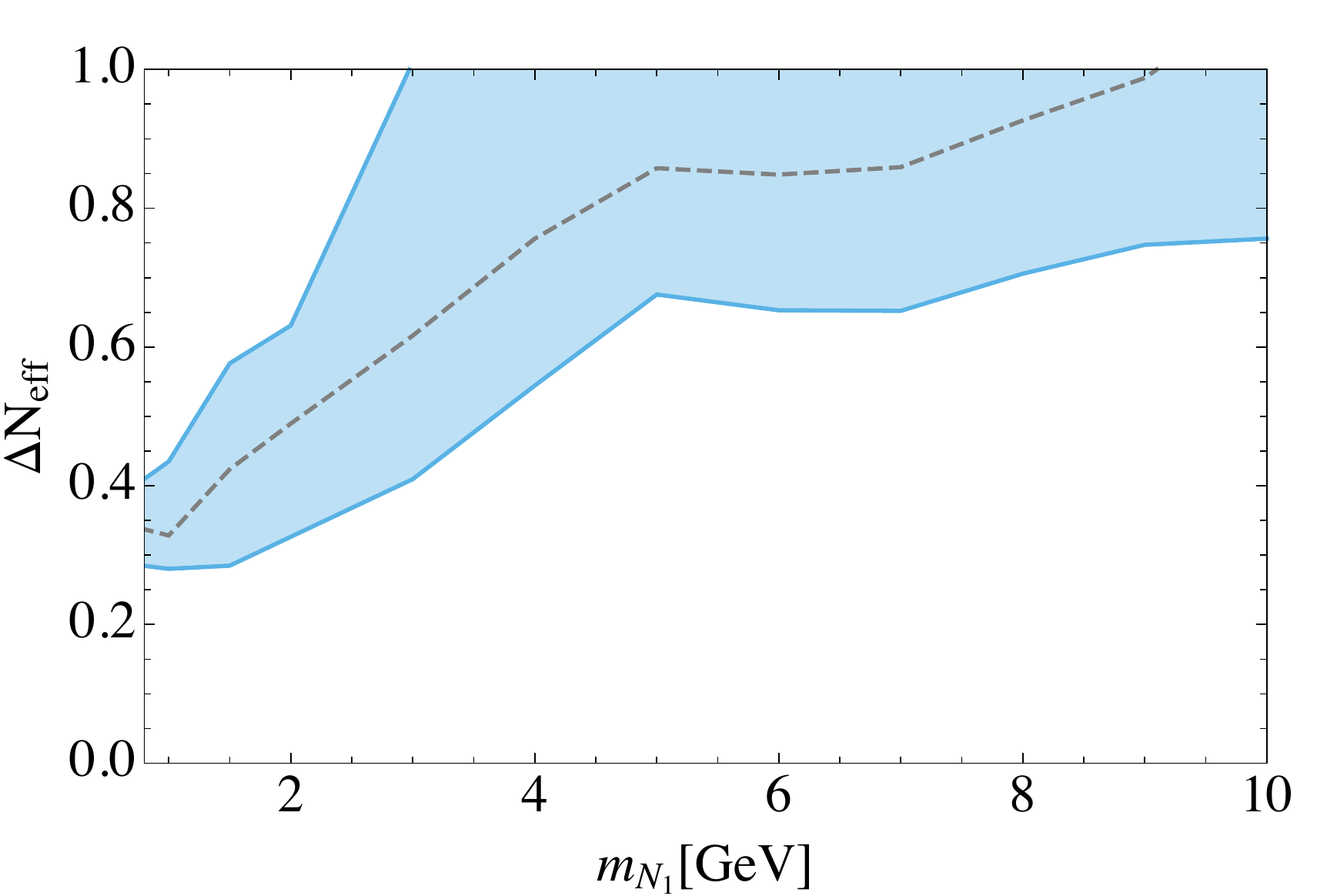} 
   \caption{
The range of $\Delta N_\mathrm{eff}$ as a function of the lightest right handed neutrino mass which is produced in an ensemble of phenomenologically viable anarchic models (see the text for details). We have taken $\epsilon=0$ and~0.05 in the left and right panels respectively. The dashed line shows the result of a particular model, chosen arbitrarily.
   }
   \label{fig:guidologyplot}
\end{figure}

Having chosen viable neutrino models we calculate the widths of the right handed neutrino states (see Appendix~\ref{app:decays}) and estimate $\Delta N_\mathrm{eff}$ following the numerical procedure described in Appendix~\ref{app:Numerical}. This procedure  takes the decay of all right handed neutrinos into account. The results are presented in figure~\ref{fig:guidologyplot} in which we show the range of $\Delta N_\mathrm{eff}$ in the ensemble of about 180 viable models found in our scan. The dashed line shows the result of a particular model, chosen arbitrarily. As expected, we see that models that have viable neutrino masses and mixings can also produce a viable cosmology.

\section{Conclusions}
\label{sec:conclusions}

In summary, we have proposed a simple solution to the cosmological 
challenges of the MTH scenario. We consider a framework in 
which there is a new weakly coupled particle species $N$ that decouples 
from the thermal bath while still relativistic. As the universe expands, 
these particles eventually become nonrelativistic and come to dominate 
the energy density of the universe, before decaying. These decays are 
assumed to occur at late times, after the SM and twin sectors have 
decoupled. Furthermore, the $N$ are assumed to decay preferentially into 
SM states rather than into twin states. The resulting energy density in 
the SM sector is then higher than in the twin sector, allowing the 
cosmological bounds on dark radiation to be satisfied.

We then consider a minimal extension of the original MTH 
that includes neutrino masses, the $\nu$MTH model, and show that it 
naturally possesses the necessary ingredients to realize these ideas. In 
the $\nu$MTH both the SM and mirror neutrinos acquire masses through the 
familiar seesaw mechanism, but with a low right-handed neutrino mass 
scale of order a few GeV. In this construction, the right-handed 
neutrinos play the role of the late-decaying species, and their 
out-of-equilibrium decays heat the SM and twin sector baths at late 
times. Since the weak gauge bosons of the SM are lighter than their twin 
counterparts, decays to SM states are preferred, with the result that 
the visible sector is left at a higher temperature than the twin sector. 
The contribution of the twin sector to the radiation density in the 
early universe is therefore suppressed, allowing the current BBN and CMB 
bounds to be satisfied. However, this effect is expected to be large 
enough to be discovered in future CMB experiments. Furthermore, the twin 
neutrinos are significantly heavier than their standard model 
counterparts, resulting in a sizable contribution to the overall mass 
density in neutrinos that can be detected in upcoming experiments 
designed to probe the large scale structure of the universe.

\acknowledgments

We thank Scott Dodelson, Seth Koren, Gordan 
Krnjaic, and Timothy Trott for illuminating discussions. ZC is supported in part by the 
National Science Foundation under grant PHY-1620074. NC is supported in part by the Department of Energy under the grant DE-SC0014129. Fermilab is 
operated by Fermi Research Alliance, LLC under Contract No. 
DE-AC02-07CH11359 with the United States Department of Energy.

\appendix

\section{Decays of the right-handed neutrinos} \label{app:decays}

\begin{figure}[t] 
   \centering
   \includegraphics[height=1.8in]{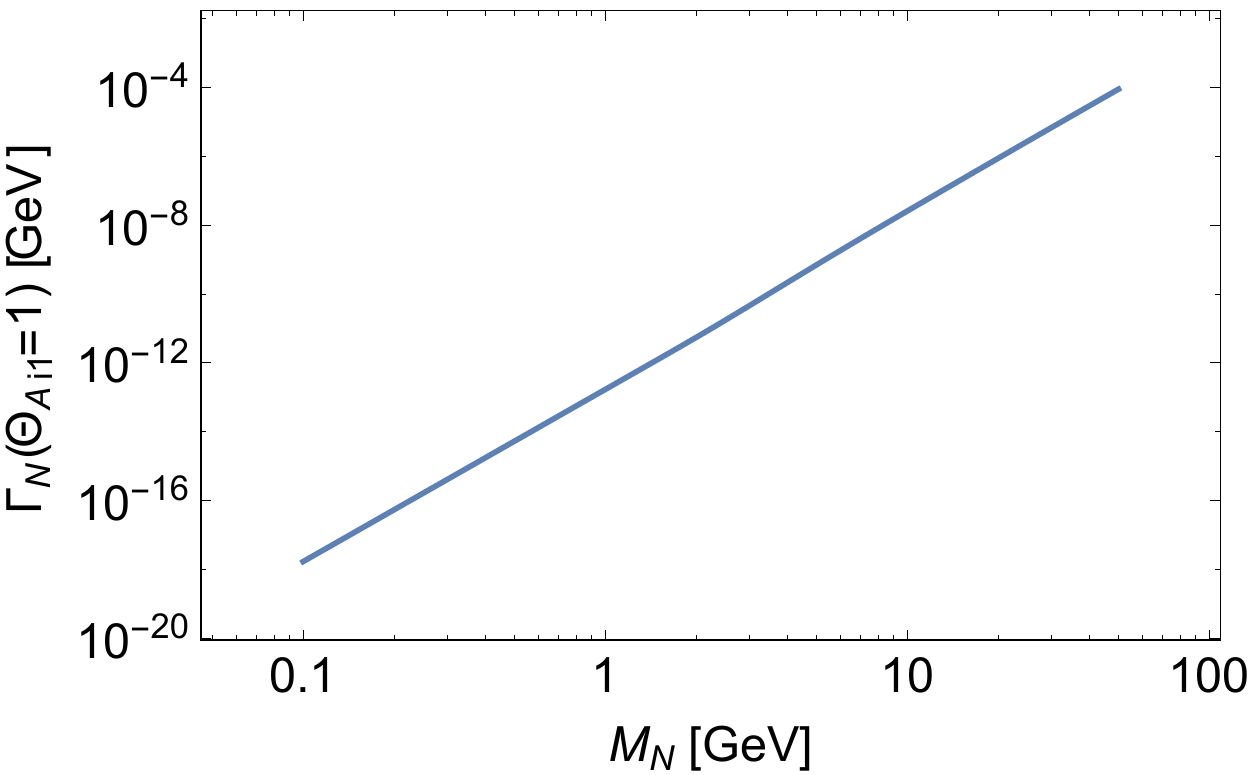} 
   \includegraphics[height=1.8in]{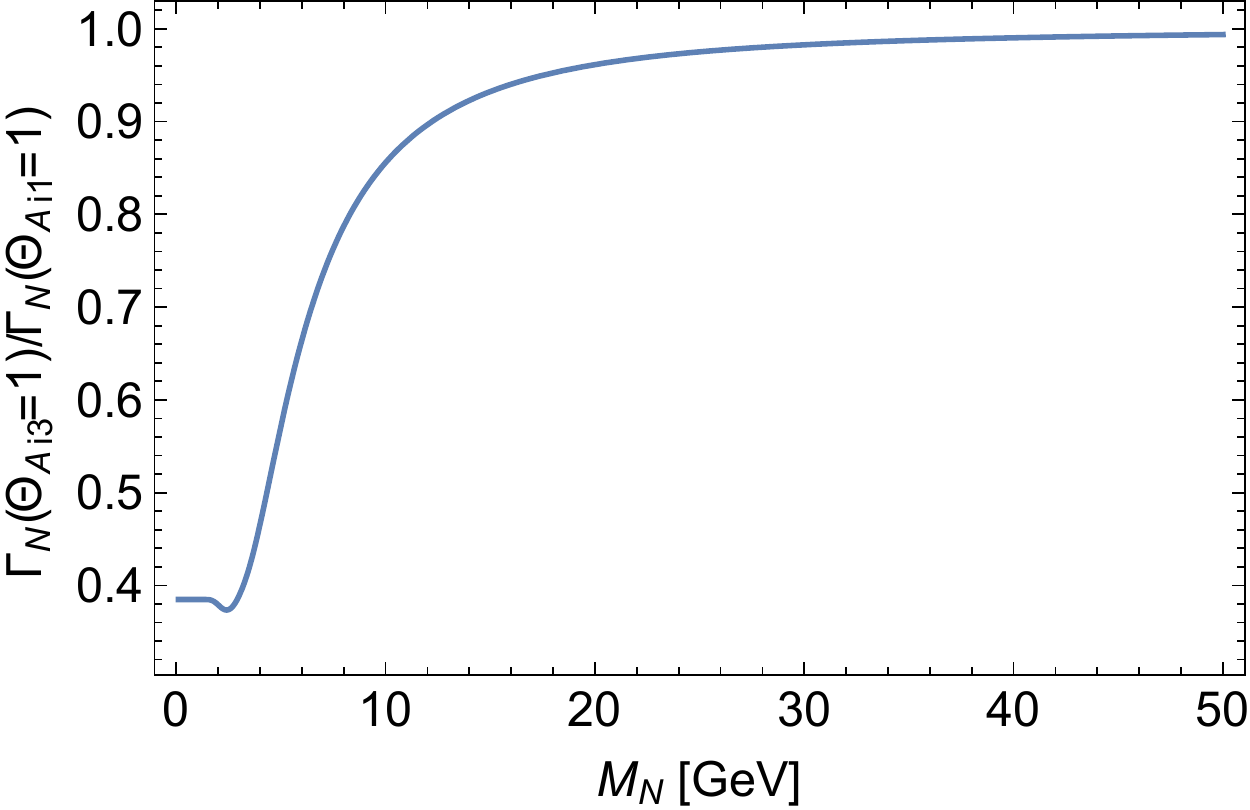} 
   \caption{Left: The partial width of right-handed neutrino $N_i$ into Standard Model states assuming $(\Theta_{A})_{i1}=1$ and all other mixing angles are zero. Right: The effect of phase space corrections due to finite bottom, charm, and tau masses, illustrated via the ratio of partial widths with $(\Theta_{A})_{i3}=1$ and $(\Theta_{A})_{i1}=1$, in each case assuming all other mixing angles are zero.}
   \label{fig:width}
\end{figure}

Decays of the right-handed neutrinos proceed through both charged and neutral weak currents, which may interfere depending on the final state. Neglecting phase space corrections due to finite quark and lepton masses, the partial widths for the decay of a right-handed neutrino $N_i$ into the Standard Model sector are given by
\begin{eqnarray*}
\Gamma(N_i \to \nu_j u_k \bar u_k) &=& N_c \frac{ |(\Theta_A)_{ij}|^2}{192 \pi^3} G_F^2 M_N^5  \left( \frac{1}{4} - \frac{2}{3} s_W^2 + \frac{8}{9} s_W^4 \right) \\
\Gamma(N_i \to \nu_j d_k \bar d_k) &=& N_c \frac{ |(\Theta_A)_{ij}|^2}{192 \pi^3} G_F^2 M_N^5  \left( \frac{1}{4} - \frac{1}{3} s_W^2 + \frac{2}{9} s_W^4 \right) \\
\Gamma(N_i \to \ell_j u_k \bar d_k) &=& N_c \frac{|(\Theta_A)_{ij}|^2}{192 \pi^3} G_F^2 M_N^5 \\
\Gamma(N_i \to \sum_k \nu_j \bar \nu_k \nu_k) &=& \frac{ |(\Theta_A)_{ij}|^2}{192 \pi^3} G_F^2 M_N^5 \\ 
\Gamma(N_i \to \ell_j \bar \ell_k \nu_{k}) &=&  \frac{ |(\Theta_A)_{ij}|^2}{192 \pi^3} G_F^2 M_N^5 \qquad (j\neq k) \\
\Gamma(N_i \to \nu_j \bar \ell_k \ell_k) &=&  \frac{ |(\Theta_A)_{ij}|^2}{192 \pi^3} G_F^2 M_N^5 \left( \frac{1}{4} - s^2_W + 2 s^4_W \right) \qquad (j\neq k) \\
\Gamma(N_i \to \ell_j \bar \ell_j \nu_j) &=&  \frac{ |(\Theta_A)_{ij}|^2}{192 \pi^3} G_F^2 M_N^5 \left( \frac{1}{4} + s^2_W + 2 s^4_W \right) 
\end{eqnarray*}
where the angles $(\Theta_A)_{ij}$ are given in Eq.~(\ref{eq:angles}) and are naturally $\mathcal{O}(\sqrt{m_{\nu,A}/M_N})$. Of course, phase space corrections to these expressions are often relevant for right-handed neutrinos in the mass range of interest; the numerical impact of nonzero bottom, charm, and tau masses is illustrated in figure \ref{fig:width}.

Expressions for decays into the twin sector may be obtained by analogy. 
The relative $\mathcal{O}(v^2/f^2)$ suppression of decays into the twin 
sector noted in Section \ref{sec:neutrinomodel} for the case in which 
the neutrino sector respects the $\mathbb{Z}_2$ twin symmetry arises from the 
combination of $G_{F,B} = \frac{v^2}{f^2} G_{F,A}$ and $\Theta_B \simeq 
\frac{f}{v} \Theta_A$.

\section{Numerical Cosmology}\label{app:Numerical}

In the analysis of Section~\ref{sec:cosmology} we presented analytic results for the effect on $N_{\mathrm{eff}}$ of late decaying sterile neutrino(s).  These results assume the sterile neutrinos decay instantaneously and that all $N$ are identical, with the number of $N$ only entering through the combination $g_{*N} M_N$ (the case of one sterile neutrino corresponds to $g_{*N}=3.5$).
However, in realistic neutrino mass models these assumptions are no longer true.  The decays of $N$ take place over an extended period and not all sterile neutrinos are the same.  In the aligned case, section~\ref{sec:universal}, all the right handed neutrinos are degenerate, up to $\mathcal{O}(M_{AB}^2/M_N)$ corrections, but their lifetimes scale as $\Gamma_i\sim m_\nu^i$ (\ref{eq:width3fam}) meaning the $N_i$ which decays to the lightest SM neutrino can be arbitrarily long lived, but we will not consider neutrino masses that cause $N$ to decay after BBN, $T\sim 1$ MeV.  In the pseudo $\mu\tau$-anarchy model, section~\ref{sec:Guidology}, there are three sterile neutrinos which can decay to all three SM neutrinos.  This anarchic mixing typically results in masses and lifetimes that can differ by up to an order of magnitude.  

Naively one might expect the result is dominated by the longest lived right-handed neutrino.  However, in general none of the decays can be ignored and instead a more careful analysis must be carried out.  As the $N$'s do not decay instantaneously, the periods over which the majority of each species decays can overlap.  When each $N$ decays it reheats the $A$ and $B$ sectors and the universe becomes less matter dominated.  Since $a_{}\sim t^{2/3}$ during matter domination whereas $a_{}\sim t^{1/2}$ during radiation domination, it takes longer for each remaining $N$ to again dominate the evolution of the universe.  Furthermore, during the decay period the $A$ and $B$ sectors are crossing various particle thresholds at different times.  Rather than attempt an analytic result we will study the evolution of the universe numerically.  

The system we wish to study is that of three massive neutrinos decaying into two baths of relativistic particles in an expanding universe, whose scale factor is $a(t)$.  Since the number of relativistic degrees of freedom changes with time as various species in each bath drop out of equilibrium, the baths are most conveniently described by the entropy of the relativistic particles in the $A$ ($B$) sector, $S_A(S_B)$.  Their evolution is governed by
\bea
\dot{S}_A&=\left(\frac{2\pi^2 g_{*A}}{45\, S_A}\right)^{1/3} a^4 \sum_i (1-\epsilon)(n_i-n_i^{eq})\langle E\, \Gamma_i\rangle 
&\approx\left(\frac{2\pi^2 g_{*A}}{45\, S_A}\right)^{1/3} a^4 \sum_i (1-\epsilon)\Gamma_i \rho_i
~, \nonumber\\
\dot{S}_B&=\left(\frac{2\pi^2 g_{*B}}{45\, S_B}\right)^{1/3} a^4 \sum_i \epsilon\, (n_i-n_i^{eq})\langle E\, \Gamma_i\rangle
&\approx  \left(\frac{2\pi^2 g_{*B}}{45\, S_B}\right)^{1/3} a^4 \sum_i \epsilon\, \Gamma_i \rho_i~.
\label{eq:cosmodiffeqs1}
\eea
In the last steps of equation~(\ref{eq:cosmodiffeqs1}) we have used the fact since that most of the decays occur when the $N_i$ are out of equilibrium and non-relativistic, the equilibrium number density can be ignored, $n_i^{eq}\approx 0$. Furthermore, the thermally averaged width is $\langle E \Gamma_i\rangle \approx M_{N_i} \Gamma_i$.  Making the same approximations the evolution of the energy densities in the sterile neutrino sector is,
\be
\dot{\rho}_i= - 3H \left(\rho_i + p_i\right) - (n_i-n_i^{eq}) \langle E\,\Gamma_i\rangle 
 \approx - 3H \left(\rho_i + p_i\right) -  \Gamma_i \rho_i~.
\label{eq:cosmodiffeqs2}
\ee
The energy density of the $A$ and $B$ sectors is related to their entropy $\rho=\frac{3}{4} (45/2\pi^2 g_*)^{1/3} S^{4/3}a^{-4}$, and the Hubble constant is $H^2=(\dot{a}/a)^2=(\rho_1+\rho_2+\rho_3+\rho_A+\rho_B)/3M_{Pl}^2$.  The pressure term, $p_i$, keeps track of the transition from $N$ behaving as radiation to matter as it cools.  Rather than keep track of the full phase space distribution we assume the average momentum of the $N_i$ just red shifts and approximate this as
\be
p_i\approx \rho/3 \times (p(t_0) a_0/a)^2/(M_N^2 + (p(t_0) a_0/a)^2)~,
\ee 
where $p(t_0)$ is the average initial momentum, and $a_0$ the scale factor, when $N$ decouples.

Although all three $N_i$ leave the bath at different temperatures (\ref{eq:Ndecouplingtemp}) these temperatures are sufficiently close that, for simplicity, we assume they all decouple at the highest decoupling temperature.  This assumption ignores the possible effect of a particle in either the $A$ or $B$ sectors leaving the bath between two decoupling temperatures and reheating those $N$'s that are still coupled.  This reheating effect, if it occurs, is small since the relative change in $g_*$ at these temperatures is not large.  Starting at the highest decoupling temperature we populate entropies and energy densities as expected for thermal distributions and then evolve according to (\ref{eq:cosmodiffeqs1}) and (\ref{eq:cosmodiffeqs2}). 

The results of solving cosmology numerically are show in figures~\ref{fig:universalplot} and \ref{fig:guidologyplot}.  We assume that the $B$ sector particles are three times heavier than their twins in the SM, \ie\ $f/v=3$.  We consider two choices for the branching ratio into the twin sector: $\epsilon=0.05$, which is slightly smaller than expected from $\mathbb{Z}_2$ alone, and the limiting case of $\epsilon=0$ where the only contribution to $\Delta N_{\mathrm{eff}}$ comes from the primeval energy density, $R_N$.


\bibliography{twincosmo}
\bibliographystyle{JHEP}

\end{document}